\documentclass[twocolumn]{aastex62}

\usepackage{graphicx}   
\usepackage{amsmath}    
\usepackage{amssymb}    
\usepackage{comment}
\usepackage{xspace}
\usepackage{lastpage}

\usepackage{eso-pic}

\usepackage{comment}

\def\reff@jnl#1{{\rm#1\/}}
\def\aj{\reff@jnl{AJ}}         
\def\araa{\reff@jnl{ARA\&A}}      
\def\apj{\reff@jnl{ApJ}}        
\def\apjl{\reff@jnl{ApJ}}        
\def\apjs{\reff@jnl{ApJS}}       
\def\aap{\reff@jnl{A\&A}}        
\def\aapr{\reff@jnl{A\&A~Rev.}}     
\def\aaps{\reff@jnl{A\&AS}}       
\def\mnras{\reff@jnl{MNRAS}}      
\def\physrep{\reff@jnl{Physics Reports}}
\def\prd{\reff@jnl{Phys.Rev.D}}     
\def\prl{\reff@jnl{Phys.Rev.Lett}}   
\def\pasp{\reff@jnl{PASP}}       
\def\pasj{\reff@jnl{PASJ}}       
\def\nat{\reff@jnl{Nature}}       
\def\jcap{\reff@jnl{JCAP}}   
\def\memsai{\reff@jnl{MemSAI}} 
\def\na{\reff@jnl{New Astronomy}}       
\def\procspie{\reff@jnl{SPIE}}       
\def\pasa{\reff@jnl{PASA}}

\defcitealias{Manolopoulou:2017}{M17}

\newcommand{\smicanosz}{\texttt{SMICA-noSZ}}

\newcommand{\planck}{{\it Planck}}
\newcommand{\howmanyclusters}{13}

\newcommand{\msol}{M_{\odot}}
\newcommand{\muK}{\mu {\rm K}}
\newcommand{\Dstat}{D}

\begin{document}

\title{
Constraining the Rotational Kinematic Sunyaev-Zel'dovich Effect in Massive Galaxy Clusters
}

\author{Eric J.~Baxter}
\affiliation{Department of Physics and Astronomy, University of Pennsylvania, Philadelphia, PA 19104, USA}
\correspondingauthor{Eric J. Baxter}
\email{ebax@sas.upenn.edu}

\author{Blake D.~Sherwin}
\affiliation{DAMTP, University of Cambridge, Cambridge CB3 0WA, UK}
\affiliation{Kavli Institute for Cosmology, Cambridge CB3 0HA, UK}

\author{Srinivasan Raghunathan}
\affiliation{Department of Physics and Astronomy, University of California, Los Angeles, CA 90095, USA}

\begin{abstract}
We constrain the rotational kinematic Sunyaev-Zel'dovich (rkSZ) effect in {\it Planck} data using a sample of  rotating galaxy clusters identified in the Sloan Digital Sky Survey (SDSS).  We extract cluster-centered cutouts from {\it Planck} cosmic microwave background (CMB) maps that have been cleaned of thermal SZ signal.  Using previous constraints on the cluster rotation vectors determined from the motions of galaxies, we fit for the amplitude of the rkSZ effect in the CMB cutouts, marginalizing over parameters describing the cluster electron distribution.  We also employ an alternative, less model-dependent approach to measuring the rkSZ signal that involves measuring the dipole induced by the rkSZ in rotation velocity-oriented CMB stacks.  In both cases, we find roughly $2\sigma$ evidence for a rkSZ signal consistent with the expected amplitude and morphology.  We comment on future directions for measurements of the rkSZ signal.  
\vspace{0.8cm}
\end{abstract}

\section{Introduction}

Most of the baryonic mass of galaxy clusters is in the form of ionized gas that makes up the intracluster medium (ICM), and which has been heated to temperatures of order $T \sim 10^7\,{\rm K}$. Characterizing the properties of this gas, including its density profile, bulk motion, and thermal state is important for our understanding of structure and galaxy formation, and for extracting cosmological constraints from observations of galaxy clusters \citep[for a review, see e.g.][]{Kravtsov:2012}. Because of its high temperature, the ICM radiates thermally in x-rays; observations of this emission are sensitive to the density and temperature of the gas, and have long been used to study and detect galaxy clusters.  

An alternate route to probing the ICM is via the Sunyaev-Zel'dovich (SZ) effect \citep{SZ_1972}, which results from inverse Compton scattering of cosmic microwave background (CMB) photons with ionized gas.  This scattering process leads to a detectable signature in submillimeter bands.  The net SZ signature of the cluster gas can be divided into two parts: one due to the thermal motion of the cluster gas (thermal SZ, or tSZ), and one due to bulk motion of the cluster gas (kinematic SZ, or kSZ).  The amplitude of the tSZ signal is sensitive to the gas pressure along the line of sight, while the amplitude of the kSZ signal is sensitive to a product of the gas density and bulk velocity relative to the CMB frame \citep[for a review, see][]{Brikinshaw:1999}.    

The kSZ effect can in turn be divided into contributions from the motion of the galaxy cluster as a whole, and contributions from internal motions of the cluster gas.  The former appears as a monopole-like temperature fluctuation centered on the cluster, and which traces the density profile of the cluster gas.  The amplitude and sign of this signal depends on how quickly the cluster is moving towards or away from the observer.    The kSZ due to internal gas motions, on the other hand, can have a complicated morphology as a result of bulk flows and turbulent motion \citep{Baldi:2018}.  Bulk rotation of the cluster gas will introduce a dipole-like kSZ signal if one side of the cluster is moving towards the observer, while the other is moving away.  We refer to this signal as the rotational kSZ, or rkSZ.

Detection of the kSZ signal is challenging because of its small amplitude, small scale, and because (unlike the tSZ) its dependence on frequency is the same as that of the primordial CMB fluctuations.  This is simply because the kSZ is effectively a Doppler boost to the CMB photons, which therefore preserves their blackbody spectrum.  Despite these observational challenges, the kSZ due to the bulk motions of halos has recently been detected in several works at roughly the $4\sigma$ level \citep[e.g.][]{Hand:2012, Soergel:2016, Hill:2016}.  Measurements of the  kSZ effect from internal motions of gas in individual clusters have also been recently reported \citep{Sayers:2013,Adam:2017}. 

In this analysis, we consider specifically the rkSZ contribution to the total kSZ signal, and how this signal correlates with the rotation of cluster member galaxies.   The rkSZ signal has been modeled analytically by \citet{Cooray:2002} and \citet{Chluba:2002}, and recently using simulations by \citet{Baldi:2018}.   The amplitude of the signal is dependent on the gas rotation velocity and density, and is expected to be of order 30~$\muK$ at peak for rapidly rotating clusters with mass  $\sim 10^{15} \msol$ \citep{Baldi:2018}.      

The rkSZ signal provides a unique means to probe the rotational component of cluster gas motion.  The total motion of gas in clusters includes coherent bulk flows --- such as that resulting from rotation --- as well as turbulent motion.  Both coherent and turbulent motion can impact observables relevant to cosmological constraints from x-ray and tSZ-selected cluster samples \citep[e.g.][]{Lau:2009}.  Especially in the inner parts of clusters, pressure support from gas rotation can be significant, motivating attempts to measure the rkSZ signal \citep{Fang:2009, Lau:2009, Lau:2013}.  Moreover, measurement of the rkSZ can be used to probe the evolution of the cluster angular momentum during structure formation \citep{Cooray:2002}.  Potentially, knowledge of the rotation of halos can be related to the density field at early times, which could have several cosmological applications (e.g. constraining dark energy or neutrinos). Finally, the rkSZ is a potentially important systematic for measurements of gravitational lensing of the CMB by galaxy clusters  \citep[e.g.][]{Lewis:2006,Baxter:2015} and for other higher-order effects in the CMB such as the moving lens effect \citep[e.g.][]{Hotinli:2018}.

In this work, we constrain for the first time the rkSZ signal correlated with the bulk rotation of cluster member galaxies.  To this end, we use maps of the CMB temperature  produced by the \planck{} satellite \citep{Planck:compsep}, and samples of rotating galaxy clusters identified in data from the Sloan Digital Sky Survey by \citet{Manolopoulou:2017}.  We use the cluster rotation velocities inferred from the galaxy velocities by \citet{Manolopoulou:2017} to inform our model for the gas rotation, a reasonable procedure since we expect the galaxy and gas rotation to be correlated \citep{Baldi:2017}.   

We develop two different methods to measure the rkSZ signal, and obtain consistent results between the two approaches.  The first approach involves constructing an explicit model for the rkSZ signal, and fitting the data to obtain constraints on the model parameters.  We focus on three parameters: one that controls the amplitude of the signal, one that controls its shape, and one that controls its maximum extent.  The second approach is more agnostic about the shape of the signal, and involves  measuring the amplitude of a dipole signal correlated with the expected rkSZ orientation.  By combining measurements from \howmanyclusters{} galaxy clusters, we find roughly $2\sigma$ evidence for the rkSZ signal using both approaches. 

The paper is organized as follows.  We present the CMB and galaxy cluster datasets in \S\ref{sec:data}. 
Our rkSZ model is described in \S\ref{sec:model}. 
We describe our analysis methodology and simulated data sets in  \S\ref{sec:methods}. 
We present the results of our measurements in \S\ref{sec:results}.  Finally, we discuss potential sources of systematic error and prospects for future measurements in \S\ref{sec:discussion}.

\section{Data}
\label{sec:data}

\subsection{CMB data}

We use the \smicanosz{} map from {\it Planck} \citep{Planck:compsep} in this analysis.  The Spectral Matching Independent Component Analysis (\texttt{SMICA}) method produces an estimated CMB map from a linear combination of multi-frequency sky maps in harmonic space \citep{Cardoso:2008}.  The linear combination chosen by \texttt{SMICA} ensures unit response to any signal with the spectral dependence of the primary CMB.  Since the kSZ (and rkSZ) has the same spectral dependence as the primary CMB, it should be preserved in the \texttt{SMICA} maps.  The resolution of the \texttt{SMICA} map is five arcminutes. 

Unlike the standard \texttt{SMICA} maps, the \smicanosz{} maps additionally impose a linear constraint to null components with the frequency dependence of the tSZ. This is important for our analysis, since the tSZ signal is large at the locations of massive galaxy clusters.  Since the tSZ is not expected to correlate with cluster rotation velocity, it likely does not constitute a significant source of bias.  However, given the large amplitude of the tSZ relative to the rkSZ, and its similar angular scale on the sky, it is an important source of noise, and the use of tSZ-nulled maps is well motivated.  

\subsection{Cluster rotation data}
\label{sec:rot_data}

\citet{Manolopoulou:2017} (hereafter \citetalias{Manolopoulou:2017}) determined the rotation amplitudes and orientations of a sample of low redshift, massive galaxy clusters using spectroscopic determinations of cluster member velocities from the Sloan Digital Sky Survey DR10 \citep{Ahn:2014}.  The \citetalias{Manolopoulou:2017} method identifies cluster rotation based on the appearance of coherent galaxy member velocities which change sign across an axis of symmetry.  The inferred symmetry axis then determines the orientation angle, $\theta_{\rm rot}$, of the cluster rotation axis.  We use the \citetalias{Manolopoulou:2017} rotation measurements in this analysis.  The cluster sample spans a redshift range from $z \sim 0.02$ to $z \sim 0.1$.  We also use the cluster richness estimates, $n_{\rm mem}$, tabulated in \citetalias{Manolopoulou:2017}.  We do not include the lowest redshift cluster in our analysis (Abel 426), since it overlaps with a masked region in the \smicanosz{} maps. 

The measurements from \citetalias{Manolopoulou:2017} do not always unambiguously identify rotation associated with a cluster.  In some cases, for instance, significant rotation may only be found in the cluster outskirts, or substructures within the cluster may not yield consistent rotation estimates.   In our fiducial analysis, we include all clusters that are identified as having significant rotation within $1.5\,h_{70}^{-1}$~Mpc, except clusters that only show significant rotation under the `loose' criterion of \citetalias{Manolopoulou:2017}.  This selection results in 13 clusters.  We also consider a more conservative selection that removes (a) clusters that do not show rotation at large radii, $R < 2.5\,h_{70}^{-1}{\rm Mpc}$, (b) clusters that have significant substructure, as determined by \citetalias{Manolopoulou:2017}, and (c) clusters that only show rotation when excluding the inner core with $R < 0.3\,h_{70}^{-1}{\rm Mpc}$.  The resultant selection consists of six clusters.  As we discuss below, the conservative selection yields similar results to the fiducial analysis, albeit with somewhat lower statistical significance.  

\section{Model} 
\label{sec:model}

Following \citet{Cooray:2002} and \citet{Chluba:2002}, we model the cluster rotation profile using a solid body rotation model.   The method of \citetalias{Manolopoulou:2017} is most sensitive to clusters whose rotation axes are oriented orthogonal to the line of sight.   For simplicity, we therefore assume that all the clusters in our sample are oriented in this way.  If this assumption is substantially incorrect for some clusters, our constraints on the amplitude of the rkSZ signal may be biased low as a non-zero inclination relative to the line of sight would reduce the signal amplitude. While such an error would complicate the interpretation of rkSZ effects, it should not induce a spurious signal.
We return to this discussion in \S\ref{sec_model_fit_results}.

We work in polar coordinates defined in the plane of the sky, $(R,\theta)$, such that $R$ is the radial separation between a point of interest in the map and the cluster center, and $\theta$ is the azimuthal angle in the plane of the sky, measured relative to the orientation of the rotation axis.  In these coordinates, the temperature signal is 
\begin{eqnarray}
\label{eq:temp}
\frac{\Delta T(R,\theta)}{T_{\rm CMB}} = \int_{-\infty}^{\infty} \, \frac{v_{||}(R, \theta, l)}{c} \sigma_T n_e(R, \theta, l) \, dl,
\end{eqnarray}
where $T_{\rm CMB} = 2.73~{\rm K}$ is the CMB temperature and $v_{||}(R,\theta,l)$ is the velocity component parallel to the line of sight at the position in the cluster specified by $(R, \theta, l)$, where $l$ represents distance along the line of sight, with the cluster at the origin.  In the solid body rotation model, with angular velocity $\omega$, we have
\begin{eqnarray}
\label{eq:velocity}
v_{||}(R,\theta,l) = \omega R \sin \theta,
\end{eqnarray}
independent of $l$.  

We follow \citet{Chluba:2002} and model the electron density, $n_e$, using a truncated isothermal $\beta$ model with $\beta = 3/2$: 
\begin{eqnarray}
\label{eq:ne}
n_e(r) = \begin{cases}
n_{e,0} \left(1 + \frac{r^2}{r_c^2} \right)^{-3\beta/2} & \text{for $r < r_{\rm max}$} \\
0 & \text{for $r > r_{\rm max}$,}
\end{cases}
\end{eqnarray}
where $n_{e,0}$, $r_c$ and $r_{\rm max}$ are model parameters.  

The temperature signal can then be written as
\begin{multline}
\label{eq:model}
\Delta T(R,\theta) = \\
T_0 \frac{ u  \sin \theta}{\sqrt{1+u^2}}  \tan^{-1}\left( \frac{\sqrt{(r_{\rm max}/r_c)^2 - u^2}}{\sqrt{1 + u^2}} \right),
\end{multline}
where
\begin{equation}
T_0 \equiv 2 T_{\rm CMB}  \sigma_T n_{e,0} \omega  r_c^2 /c, 
\end{equation}
and we have defined $u = R/r_c$, and $u_{\rm max} = R/r_{\rm max}$.  To reduce degeneracy between the model amplitude and $r_c$, we define
\begin{eqnarray}
A &\equiv& n_{e,0} r_c^2 \\
&=& 9.52\times 10^{44}\,{\rm cm}^{-1} \left( \frac{n_{e,0}}{10^{-2} \,{\rm cm}^{-3}}\right)\left( \frac{r_c}{0.1\,{\rm Mpc}}\right)^2.
\end{eqnarray}
Substituting, we have
\begin{equation}
T_0 = \frac{2 T_{\rm CMB} \sigma_T}{c} A \omega.
\end{equation}

Larger clusters will have larger electron densities, larger $r_c$ and larger $r_{\rm max}$.  Very roughly, we expect $n_{e,0} \propto M$, where $M$ is the cluster mass.  Moreover, we expect (again, roughly) $M \propto n_{\rm mem}$, where $n_{\rm mem}$ is the cluster richness.  Additionally, we expect the distances describing the size of cluster --- i.e. the parameters $r_c$ and $r_{\rm max}$ --- to scale roughly with $M^{1/3}$.  We therefore have $A = n_{e,0} r_c^2 \propto n_{\rm mem}^{5/3}$.  To account for the scaling of $A$, $r_c$ and $r_{\rm max}$ with $n_{\rm mem}$, we adopt the following relations:
\begin{eqnarray}
\label{eq:An}
A = 9.52\times 10^{44}\,{\rm cm}^{-1}  A_n (n_{\rm mem}/200)^{5/3},
\end{eqnarray}
\begin{equation}
\label{eq:Arc}
r_{c} = 0.1\,{\rm Mpc}\,A_{\rm rc} (n_{\rm mem}/200)^{1/3},
\end{equation}
and
\begin{equation}
\label{eq:Armax}
r_{\rm max} = 1.0\,{\rm Mpc}\,A_{\rm rmax} (n_{\rm mem}/200)^{1/3}.
\end{equation}
The parameters $A_n$, $A_{\rm rc}$, $A_{\rm rmax}$ and $\omega$ completely specify our model, given the measured $n_{\rm mem}$.  The fiducial values at $n_{\rm mem} = 200$ are chosen to give an rkSZ signal for which the maximum amplitude is in rough agreement with the results of \citet{Baldi:2018}.

\begin{figure*}
\centering
\includegraphics[scale=0.45]{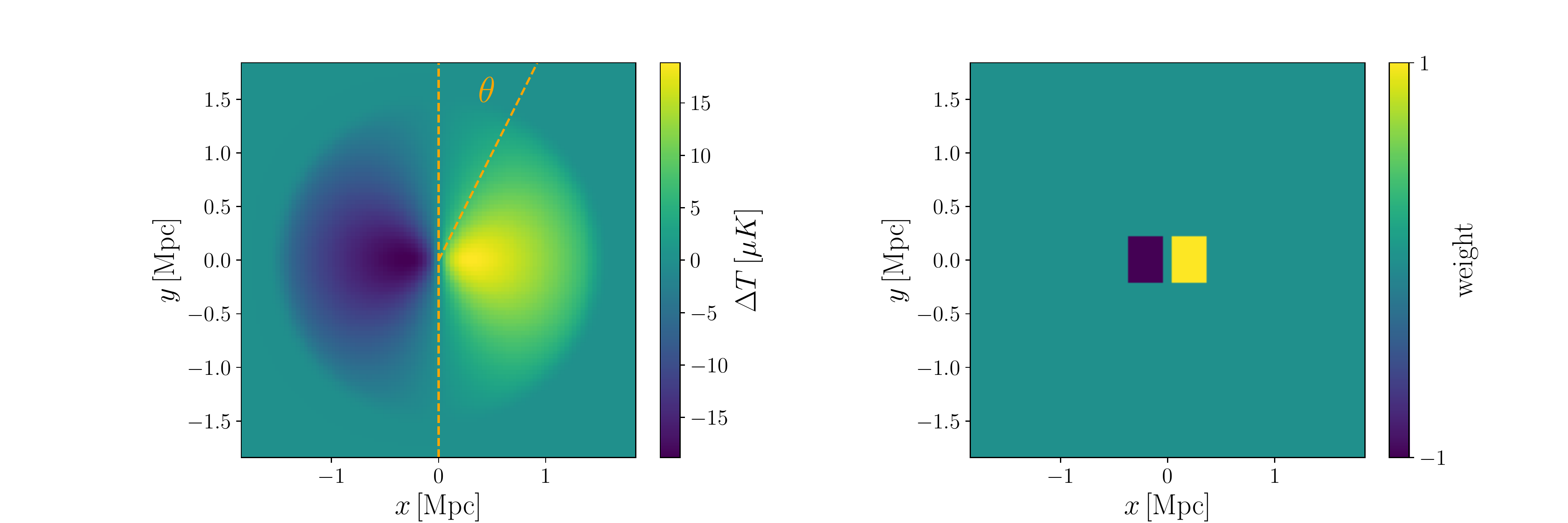}
\caption{\label{fig:model} Left: the model cluster profile, as described in \S\ref{sec:methods}.  The cluster is oriented with its rotation axis in the vertical direction (vertical orange dashed line), and is rotating such that its right side is coming out of the page. Right: the weighting applied to the cutouts for the asymmetry analysis described in \S\ref{sec:asymmetry}.  Since the CMB has power on large scales compared to the cluster size, and since the signal is expected to decline quickly away from the cluster center, we limit the asymmetry measurement to patches near the cluster center. }
\end{figure*}

Finally, we account for the beam of the \smicanosz{} maps by convolving the model profile with a Gaussian beam.  We assume $\theta_{\rm FWHM} = 5'$.  An example rkSZ model, and an illustration of our coordinate system, are shown in the left panel of Fig.~\ref{fig:model}.

\section{Methods}
\label{sec:methods}

\subsection{Temperature signals at cluster locations}

We extract patches (cutouts) from the \smicanosz{} maps centered at the locations of the clusters described in \S\ref{sec:rot_data}.  Each cutout is oriented so that the $y$-axis coincides with the direction on the sky specified by the orientation angles, $\theta_{\rm rot}$, determined by \citetalias{Manolopoulou:2017}.  The side with $x > 0$ then corresponds to part of the cluster that is moving towards the observer, while the side with $x < 0$ corresponds to the part that is moving away. 

Because the clusters have different redshifts, their rkSZ signals will necessarily span different angular scales on the sky.  To reduce this variation, we scale the cutouts to physical coordinates by interpolating them onto a grid with dimension $7.4\,{\rm Mpc}$ on a side (removing pixels outside of this range).  When we analyze the cutouts using a model fitting approach (described in \S\ref{sec:likelihood}), however, we do not interpolate onto physical coordinates, since in this case the {\it model} can be adjusted to account for differing cluster redshifts.  For the model fitting analysis, the cutouts have dimension $1^{\circ} \times 1^{\circ}$, with $1.5' \times 1.5'$ pixels.  

We now describe the two approaches we take to measuring the rkSZ signal from the cluster cutouts.

\subsection{Simple method: asymmetry analysis}
\label{sec:asymmetry}

The rkSZ induces a dipole-like signal in the CMB at the location of clusters that is oriented orthogonal to the cluster rotation axis, as seen in Fig.~\ref{fig:model}.  We first attempt to detect this asymmetric signal using a simple approach that is agnostic about its amplitude or precise shape.

We define a statistic, $\Dstat$, via
\begin{eqnarray}
\Dstat = \langle T_{\rm right} \rangle_{p} - \langle T_{\rm left} \rangle_{p},
\end{eqnarray}
where $\langle T_{\rm right} \rangle_{p}$ ($\langle T_{\rm left} \rangle_p$) is the mean of pixels to the right (left) of the rotation axis, with the subscript $p$ indicating that the mean is over different pixels.  The rkSZ signal is expected to peak at about $0.1 r_{\rm vir}$, where $r_{\rm vir}$ is the virial radius of the cluster \citep{Baldi:2018}.  For the massive clusters considered here, the virial radius is roughly $2\,{\rm Mpc}$, so we expect the signal to peak at about $\sim 0.2\,{\rm Mpc}$ from the rotation axis.  We therefore use pixels near $\pm 0.2\,{\rm Mpc}$ to measure $D$.  The precise selection of pixels used to compute $D$ is illustrated in the right panel of Fig.~\ref{fig:model}.  The selection of pixels with positive weight extends from 0.04 to 0.4 Mpc, and has a height of 0.4 Mpc; the selection with negative weight is mirrored across $x=0$.  We could in principle use a larger pixelated area to compute $D$; however, doing so would also increase the noise in the measurements from primordial CMB fluctuations.  

We define $\langle \Dstat \rangle_{c}$ as the average value of $\Dstat$ across all clusters, weighted by the cluster velocities and richnesses: 
\begin{eqnarray}
\langle \Dstat \rangle_c = \frac{\sum_{i=0}^{N_c} v_{\rm rot,i} n_{\rm mem, i} \Dstat_i }{\sum_{i=0}^{N_c} v_{\rm rot,i} n_{\rm mem,i}}.
\end{eqnarray}
This weighting is motivated since the rkSZ signal is proportional to $v_{\rm rot}$ and $n_e$, and $n_e$ should scale roughly linearly with cluster richness.

\begin{table*}
\centering
\begin{tabular}{|c|c|c|}
\hline
     Parameter &  Description  & Prior \\ \hline
     $A_n$ & Controls amplitude of rkSZ signal via Eq.~\ref{eq:An} & Flat \\ \hline 
     $A_{rc}$ & Controls shape of rkSZ signal via Eq.~\ref{eq:Arc} & Flat \\ \hline 
     $A_{\rm rmax}$ &  Controls maximum extent of rkSZ signal via Eq.~\ref{eq:Armax} & $\delta$ function \\ \hline
     $w$ & Angular velocity of the halo  & Gaussian prior from \citetalias{Manolopoulou:2017} (see Eq.~\ref{eq:wprior}) \\ \hline
\end{tabular}
\caption{\label{tab:params}The parameters varied in our analysis, and the corresponding priors.  The parameters $A_n$, $A_{rc}$ and $A_{\rm rmax}$ are global parameters describing all clusters in the sample, while $w$ is allowed to vary for each cluster.  }
\end{table*}

\subsection{Fiducial method: model fitting}
\label{sec:likelihood}

The approach to measuring the rkSZ described in \S\ref{sec:asymmetry} is fairly agnostic about the precise shape of the signal. We also consider a different approach in which we use the estimated profile from Eq.~\ref{eq:model} to constrain the rkSZ signal.  This approach has the advantage that it uses more of our knowledge about the expected signal, but has the disadvantage that it is more sensitive to our modeling assumptions.

We assume a Gaussian likelihood for the data, $\vec{d}_i$, in the $i^{\rm th}$ cutout:
\begin{eqnarray}
\label{eq:likelihood}
\ln \mathcal{L}(\vec{d}_i | \vec{\Delta T}_i) = -\frac{1}{2}(\vec{d}_i - \vec{\Delta T}_i)^T \mathbf{\hat{C}}_i^{-1} (\vec{d}_i - \Delta\vec{T}_i), 
\end{eqnarray} 
where $\vec{\Delta T}_i \equiv \Delta T(\vec{R},\vec{\theta})$ is the rkSZ model for the $i$th cluster from Eq.~\ref{eq:model}.  The vector indices here represent the fact that we measure a grid of temperature values across a single cutout.  The model profile is a function of $A_n$, 
$A_{\rm rc}$, $A_{\rm rmax}$, $\omega$ and $n_{\rm mem}$.  We use the measured $n_{\rm mem}$ for each cluster.  We will return to the estimate of $\omega$ in a moment.

We estimate the covariance matrix for each cutout, $\mathbf{\hat{C}}_i$, using the \smicanosz{} maps.  The pixel-space covariance is related to the power spectra of the maps, $C_i^{\rm tot}(\ell)$, in the vicinity of the $i^{\rm th}$ cutout via \citep[e.g.][]{Dodelson:book}
\begin{eqnarray}
\label{eq:covariance}
[\mathbf{\hat{C}}_i ]_{ab} = \int \frac{d\ell \,\ell}{2\pi}J_{0}(\ell \theta_{ab})C_i^{\rm tot}(\ell),
\end{eqnarray}
where $a$ and $b$ represent pixel indices, to be distinguished from the cutout index $i$, $\theta_{ab}$ represents the angular separation between pixels $a$ and $b$, and $J_0$ is the zeroth order Bessel function of the first kind.  We compute the power spectra of the \smicanosz{} maps in $10^{\circ} \times 10^{\circ}$ patches centered on the cluster locations in order to capture local variations in the {\it Planck} noise.   Each cluster is  analyzed using the appropriate covariance matrix, as in Eq.~\ref{eq:likelihood}.   The estimate of the pixel-space covariance in Eq.~\ref{eq:covariance} does not capture  contributions that are correlated with the clusters (other than the large scale variations in noise).  However, as we discuss in  \S\ref{sec:systematics}, such contributions are not expected to have a significant impact on our likelihood analysis.  

We use the cluster rotation measurements from \citetalias{Manolopoulou:2017} to obtain priors on $\omega$ for each cluster.  We use $v_{{\rm rot},i}$ to refer to the \citetalias{Manolopoulou:2017} velocity estimate for the $i^{\rm th}$ cluster, and use $\sigma_{{\rm vrot},i}$ to refer to the uncertainty on this quantity reported by \citetalias{Manolopoulou:2017}.  The rotation velocity is measured by \citetalias{Manolopoulou:2017} below 1.5 Mpc from the cluster center.  We will assume for simplicity that these measurements correspond to the line of sight galaxy velocity at 0.75 Mpc from the cluster center; changing this assumption does not impact the detection significance, but rather impacts the recovered values of $A_n$.  We emphasize that our primary goal is to determine whether there is evidence for an rkSZ signal correlated with the galaxy member rotation, and not to infer precise values of the model parameters.  For the $i^{\rm th}$ cutout, we adopt a Gaussian prior on $\omega$:
\begin{equation}
\label{eq:wprior}
Pr_i(\omega) = \frac{1}{\sqrt{2\pi \sigma_{\rm \omega}^2}} 
\exp \left[-\frac{\left( \omega - \bar{\omega}_i \right)^2}{2\sigma_{\omega,i}^2} \right],
\end{equation}
with $\bar{\omega}_i = v_{{\rm rot}, i}/(0.75 \, {\rm Mpc})$ and $\sigma_{\omega,i} = \sigma_{{\rm vrot}, i}/(0.75 \, {\rm Mpc})$.

For the model fits, we impose a non-informative, flat prior on $A_n$, allowing $A_n \in [-100,400]$.  We also allow $A_{\rm rc}$ to vary over a wide range with a flat prior ($A_{\rm rc} \in [0.01, 20]$).  However, given the strong degeneracy between $A_n$ and $A_{\rm rc}$, we will ultimately impose an informative (flat) prior on $A_{\rm rc}$ when computing the detection significance.  The fits of \citet{Baldi:2018} suggest that a plausible value for $r_c$ is about 0.1$r_{\rm vir}$. Given the large masses of the clusters in our sample, we impose the prior $A_{rc} \in [1.5,2.5]$, corresponding to $r_c \in  [0.15,0.25]\,{\rm Mpc}$.  For simplicity, we will keep $r_{\rm max}$ fixed in our analysis, but will explore different choices of this parameter below.  We summarize the parameters varied in the analysis and the corresponding priors in Table~\ref{tab:params}.

The posterior on the parameters from all clusters is then given by
\begin{multline}
P(A_n, A_{\rm rc}, A_{\rm rmax} |\{\vec{d}_i \} ) \propto \\
\prod_i^{N} \int d \omega \,\mathcal{L}(\vec{d}_i | \Delta\vec{T}_i(A_n, A_{\rm rc} , A_{\rm rmax}, \omega)) Pr_i(\omega),
\end{multline}
where the product runs over all $N$ clusters in the sample.

\subsection{Simulated cluster cutouts}

For the purposes of validating the methodology described above, we generate simulated observations of the rkSZ signal. Mock cluster rkSZ profiles constructed from the model described in \S\ref{sec:model} are assigned random orientations and positions on the sky, and added to the {\it Planck} \smicanosz{} maps.  This procedure will accurately capture the real noise and astrophysical background in these maps, but will not capture any backgrounds correlated with the cluster positions; we comment on this shortcoming in \S\ref{sec:discussion}.  The redshifts of the simulated clusters are drawn from the redshift distribution of the true clusters.  

Each mock cluster has its axis of rotation oriented orthogonal to the line-of-sight, and model parameters $A_{n} = 60$, $A_{rc} = 2.0$, $A_{\rm rmax} = 2.0$, $v_{\rm rot} = 300\,{\rm km}/{\rm s}$ and $n_{\rm mem} = 100$.  These choices yield a rkSZ signal with peak amplitude  $\sim 30\,\mu {\rm K}$, which is comparable to the peak amplitude seen in the simulations of \citet{Baldi:2018}.  We assume that the velocity errors for the simulated cluster measurements are only $1\,{\rm km}/{\rm s}$, significantly better than the uncertainty in the data.  This choice allows us to use fewer clusters to validate our analysis pipelines.  We emphasize that the simulations are {\it not} used to estimate errorbars for our measurements, but rather to validate the methodology described in \S\ref{sec:asymmetry} and \S\ref{sec:likelihood}.

\section{Results}
\label{sec:results}

\subsection{Stacking and asymmetry results}

\begin{figure*}
\centering
\includegraphics[scale=0.5]{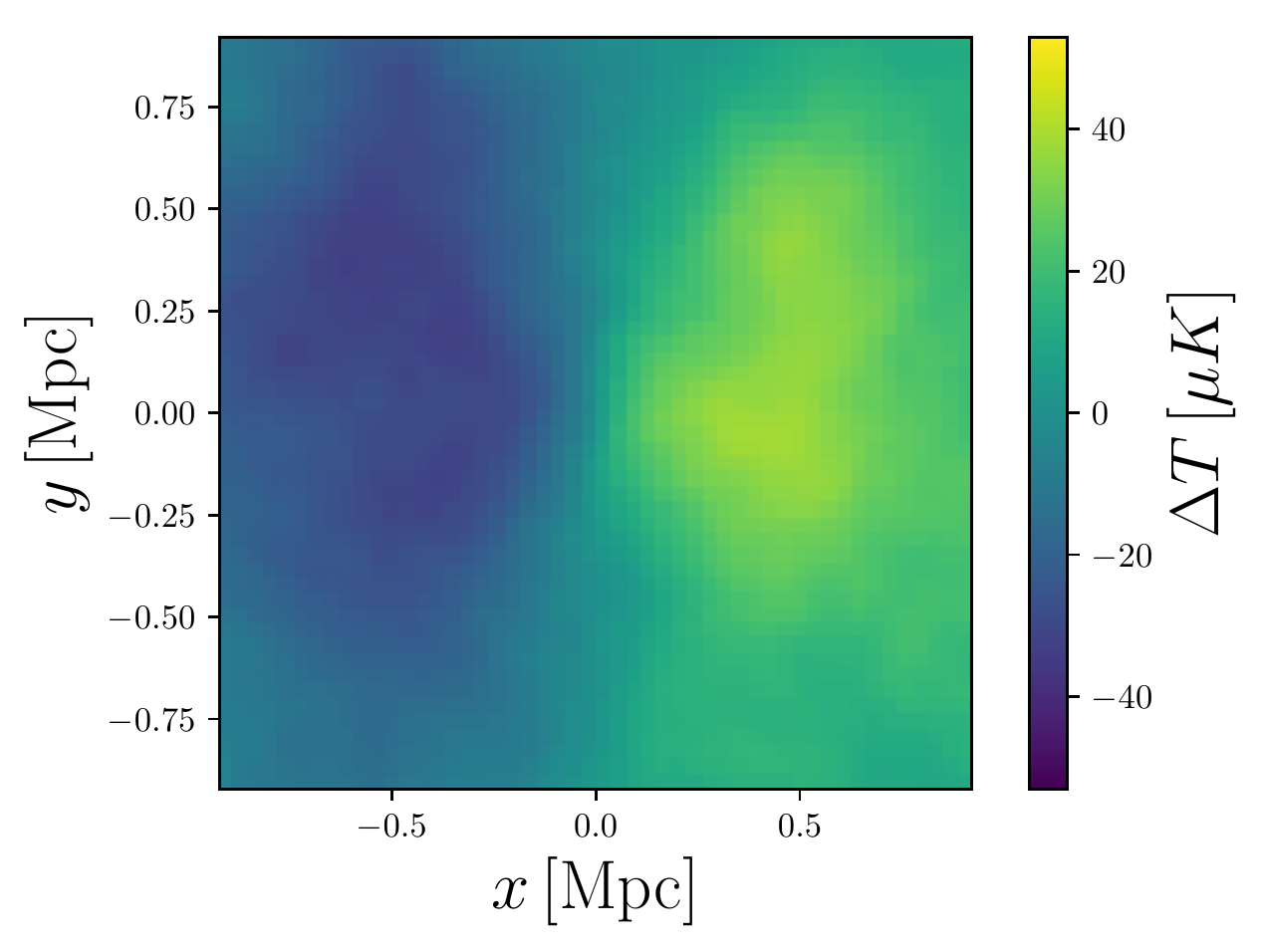}
\includegraphics[scale=0.5]{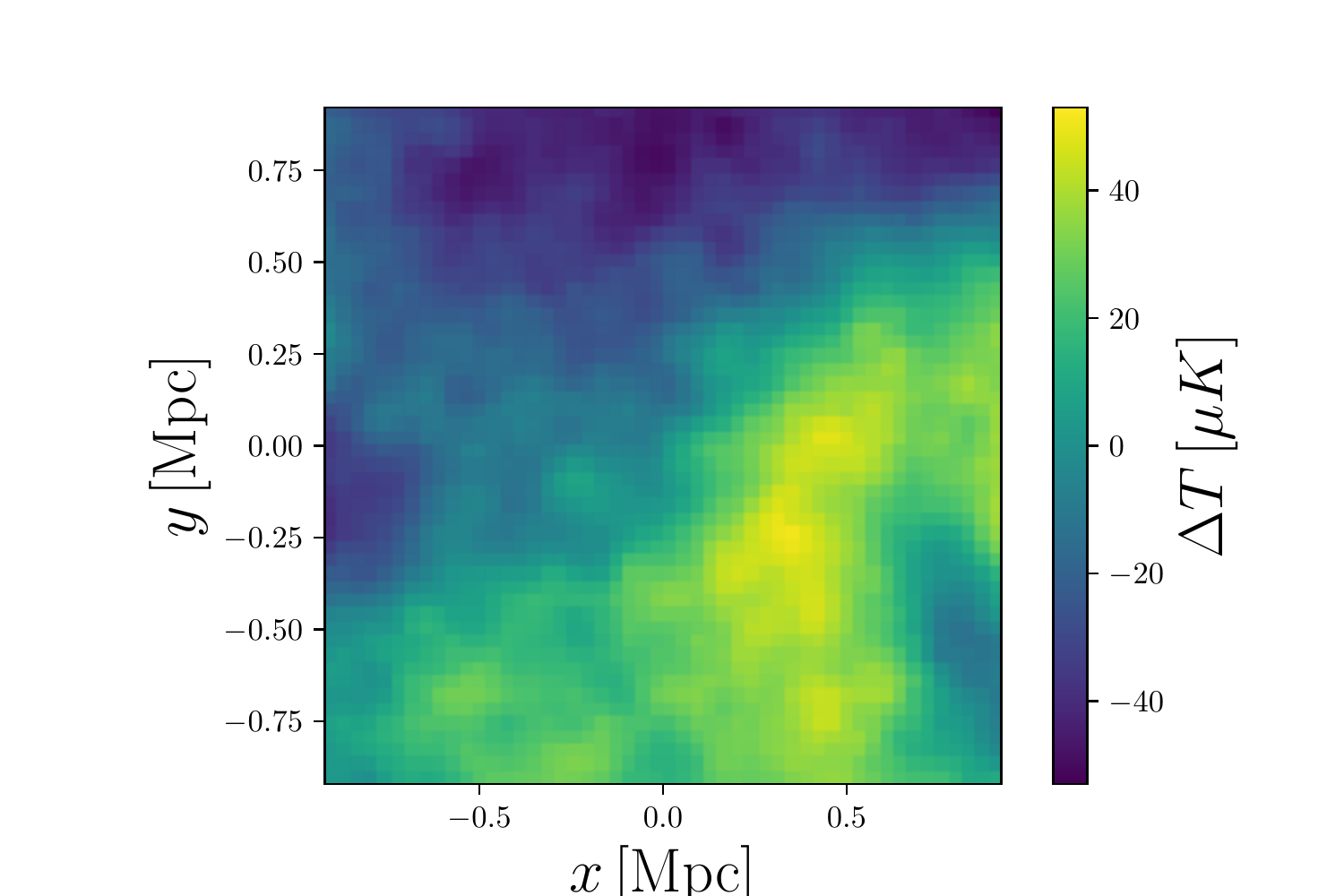}
\caption{\label{fig:weight_stack} Left: oriented, weighted stack of 100 simulated cluster cutouts that include a simulated rkSZ signal as well as real {\it Planck} noise and backgrounds.  Cutouts are oriented with the rotation axis in the vertical direction (so that the right side of the cutout is moving towards the observer), and are scaled into physical coordinates.  A clear dipole signal is seen in the stack.  Right: oriented and weighted stack of 13 cutouts from \smicanosz{} maps centered on the clusters described in \S\ref{sec:rot_data}.  There is an apparent dipole signal in the cluster cutouts that is oriented in the direction expected.  There are, however, large temperature fluctuations in the stacked map.   }
\end{figure*}

The stack of the 100 simulated cutouts is shown in Fig.~\ref{fig:weight_stack}.  There is a clear dipole signal visible at the center of the stacked cutout, despite the noise in the maps.  The stack for the 13 actual clusters is shown in the right panel of Fig.~\ref{fig:weight_stack}.  In this case, there again appears to be a dipole signal with the expected orientation.  Of course, the noise fluctuations in the stack of 13 cutouts are larger than in the stack of 100 simulated cutouts. 

From the actual cluster cutouts, we measure $\langle \Dstat \rangle_{c} = 27.6\,\mu{\rm K}$.   To assess whether this measurement is statistically significant, we also measure $\Dstat$ using sets of random points on the sky.  Each random point is assigned a random orientation, and a redshift drawn from the distribution of real cluster redshifts.  We draw a number of random points equal to the number of real clusters, and compute the average $D$ for these random points, which we call $\langle \Dstat \rangle_r$.  We then repeat this process many times to build up a distribution of $\langle \Dstat \rangle_r$.  We find that 4\% of the random point measurements have $\langle \Dstat \rangle_r > \langle \Dstat \rangle_{c}$, equivalent to a roughly $2\sigma$ measurement of an asymmetry oriented along the cluster rotation axis.  We have also varied the selection of pixels used to compute $\langle \Dstat \rangle_c$, finding that the detection significance is not sensitive to small variations in the pixel weighting.  

The maximum amplitude of the rkSZ signal seen in the simulations of \citet{Baldi:2018} is $\sim 30\,\mu{\rm K}$, corresponding to a peak-to-trough temperature difference of about $\sim 60\,\mu{\rm K}$.  Our measurement of $D$ is about a factor of two below the simulated peak-to-trough difference.  Note, though, that the measurement of $D$ is not expected to recover the full peak-to-trough difference, since the measurement of $D$ averages over regions away from the peak and trough of the signal.  Furthermore, if the rotation axes of some of the clusters in our sample are not oriented exactly orthogonal to the line of sight, that would also reduce the amplitude of the signal relative to expectations from \citet{Baldi:2018}.

The use of random points to estimate the distribution of $\langle \Dstat \rangle_c$ under the null hypothesis of no rkSZ misses any noise sources that are correlated with the clusters.  Various signals, such as the bulk kSZ signal and emission from galaxies in the clusters, are expected to correlate with the cluster positions.  However, to contribute noise to the measurement of $\langle \Dstat \rangle$, such signals must be asymmetric across the cluster rotation axes.  Any such noise sources are likely subdominant relative the contributions from noise sources that are uncorrelated with the clusters (such as primary CMB and instrumental noise).  We discuss cluster-correlated noise in more detail in \S\ref{sec:systematics}.

\begin{figure*}
\centering
\includegraphics[scale=0.5]{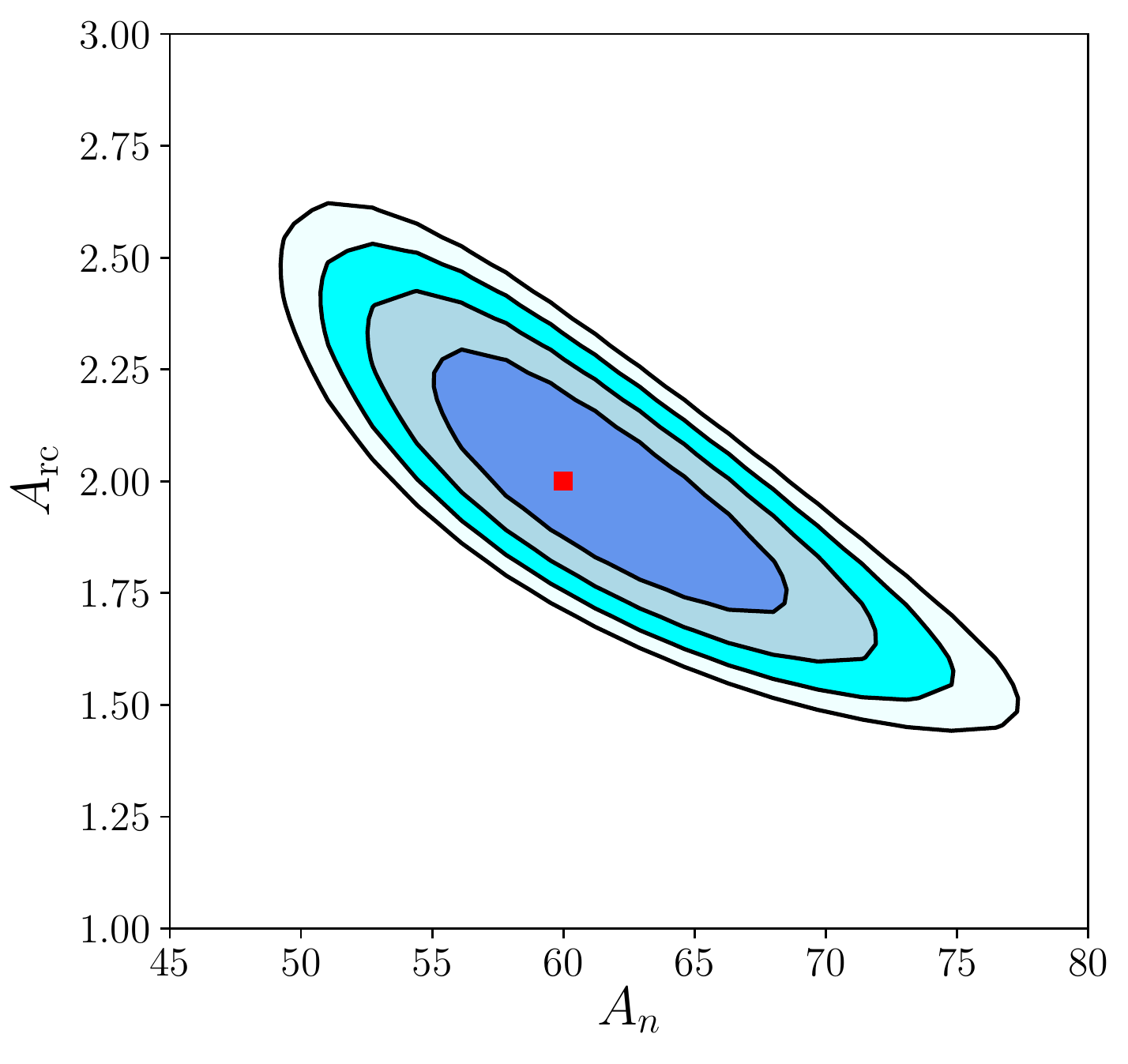}
\caption{\label{fig:likelihood_sim_results} Likelihood evaluated on 100 simulated rkSZ cutouts with {\it Planck} noise.  The parameter $A_n$ describes the amplitude of the signal, while $A_{\rm rc}$ describes its shape.  The red point indicates the input parameter choices; the analysis on simulated data recovers the input rkSZ parameters to within the errors, as expected. Contour lines represent $\Delta \chi^2 = 1$. }
\end{figure*}

\begin{figure*}
\centering
\includegraphics[scale=0.5]{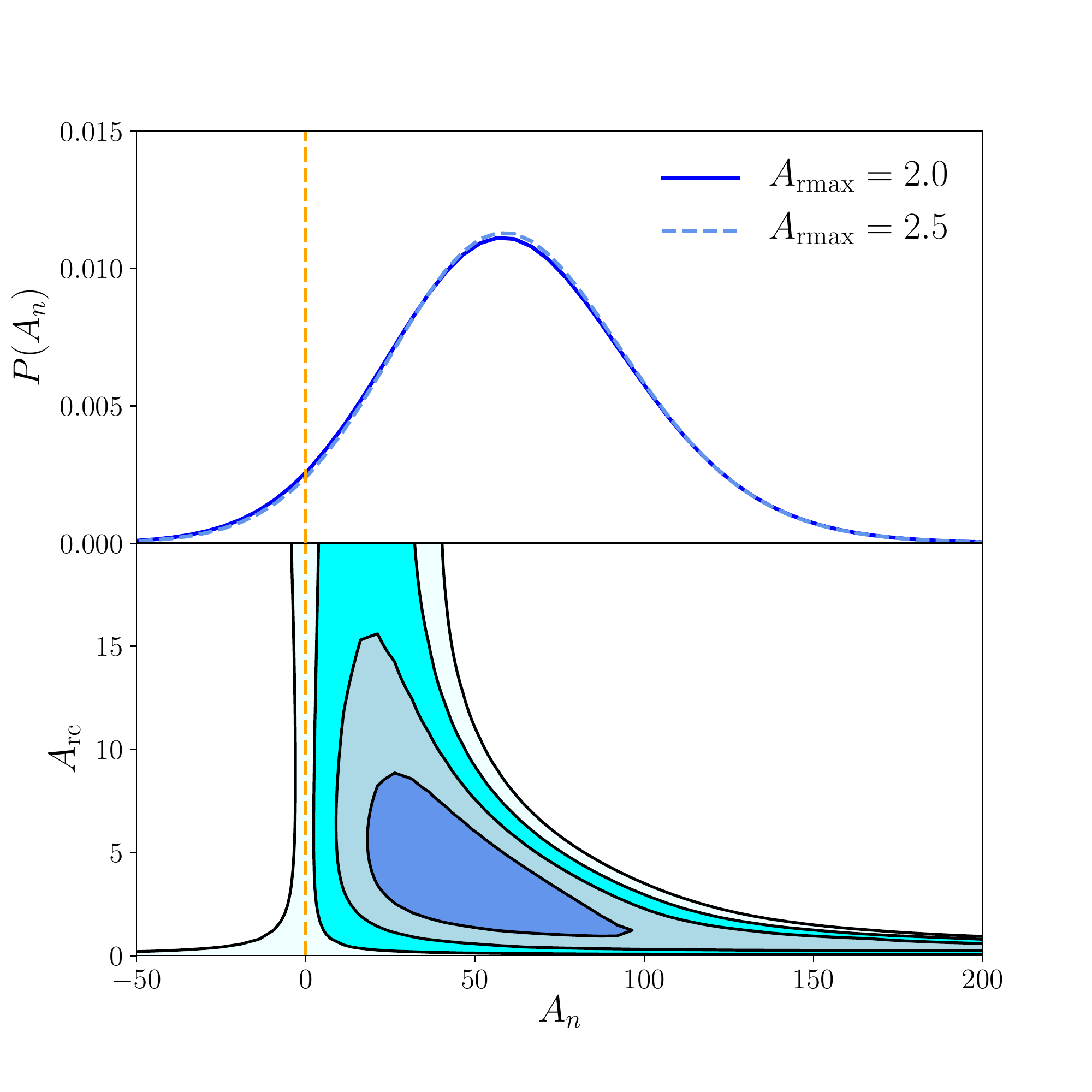}
\caption{\label{fig:likelihood_data_results} Constraints on rkSZ parameters from analysis of 13 galaxy clusters using the {\it Planck} maps.  Bottom panel shows constraints in space of $A_{n}$ (amplitude) and $A_{\rm rc}$ (shape).  Contour lines represent $\Delta \chi^2 = 1$.  Top panel shows marginalized posterior on $A_n$ after imposing an $A_{\rm rc} \in [1.5, 2.5]$ prior.  We find a preference for positive rkSZ signal at $2.1\sigma$, with a weak dependence on $A_{\rm rmax}$.  }
\end{figure*}

\subsection{Model fit results}
\label{sec_model_fit_results}
The results of the likelihood analysis applied to the 100 simulated cluster cutouts are shown in Fig.~\ref{fig:likelihood_sim_results}.   The likelihood analysis correctly recovers the input parameters to within the errorbars, as indicated by the red point in the figure.  Note that the errorbars shown in Fig.~\ref{fig:likelihood_sim_results} are for the combined analysis of 100 simulated cutouts.  There is some degeneracy between the parameters $A_n$ and $A_{\rm rc}$.

The results of the model fitting applied to data are shown in Fig.~\ref{fig:likelihood_data_results}.  In the bottom panel we show the two-dimensional posterior on $A_{n}$ and $A_{\rm rc}$.  In the top panel, we show the marginalized posterior on $A_{n}$ (imposing the $A_{\rm rc} \in [1.5, 2.5]$ prior discussed in \S\ref{sec:methods}).  We also show the result of varying $A_{\rm rmax}$.  Marginalizing over $A_{\rm rc}$ , we find $A > 0$ at $2.1\sigma$ confidence, with a weak dependence on $r_{\rm rmax}$.  The data prefer a value of $A_n$ in the range of about 20 to 70, with a corresponding temperature range of about 10 to 40~$\mu {\rm K}$.  This amplitude is consistent with the predictions of \citet{Baldi:2018}.  Our results imply an upper limit to the average rkSZ signal of 42~$\mu{\rm K}$ (84th percentile).

As noted in \S\ref{sec:model}, we have assumed that the cluster rotation vectors are oriented orthogonal to the line of sight.  Any inclination of the rotation axis relative to the line of sight would decrease the rkSZ signal, causing us to infer a low amplitude rkSZ signal.  Naively, then, one might then be surprised that the amplitude constraints are consistent with those reported by \citet{Baldi:2018}.  However, as noted in \S\ref{sec:rot_data}, the method of \citetalias{Manolopoulou:2017} is expected to be most sensitive to clusters that {\it are} oriented orthogonal to the line of sight, so agreement between our amplitude measurements and those of \citet{Baldi:2018} is not particularly surprising. 

As we have noted above, the precise values of the parameters obtained in this analysis should be interpreted with some caution, given their sensitivity to the assumptions we have made.  However, both the asymmetry analysis and the model fitting analysis provide evidence for a signal in the {\it Planck} maps that is consistent with expectations for an rkSZ signal, and which is aligned with the direction of cluster member rotation.

\section{Discussion}
\label{sec:discussion}

We have presented a constraint on the rkSZ signal from massive galaxy clusters, assuming that the direction of gas rotation for these objects is correlated with the direction of rotation of their member galaxies.  This assumption is motivated by simulation studies from e.g. \citet{Baldi:2017}.  We find roughly $2\sigma$ evidence for a rkSZ signal with approximately the expected amplitude and morphology.  Our measurement is not highly statistically significant, so we caution against over-interpretation.  However, assuming that the measurement is not a statistical fluctuation, it implies the existence of coherent gas rotation that is correlated with galaxy motion for the clusters in our sample.  

\subsection{Potential sources of systematic error}
\label{sec:systematics}

The rotational kSZ signal is unique in several respects compared to nearly all potential background noise sources.   For one, it is a dipole signal.  Most signals correlated with clusters in the \smicanosz{} maps, such as emission from dusty sources correlated with the cluster, are not expected to exhibit a dipole pattern.  

Secondly, the rkSZ signal is nearly unique among cluster-correlated, dipole-like signals in the CMB in that it is correlated with the direction of cluster rotation.  For instance, gravitational lensing of the CMB by galaxy clusters is expected to produce a dipole-like signal at the locations of galaxy clusters (see discussion of predicted signal in e.g. \citealt{Seljak:2000} and measurement in data in e.g. \citealt{Baxter:2015}).  The lensing-induced dipole, however, will be aligned with the gradient in the unlensed CMB, rather than with the orientation of the cluster rotation. 

Another dipole signal expected in the \smicanosz{} maps at the locations of galaxy clusters is the moving lens signal, discussed in \citet{Lewis:2006} and \citet{Hotinli:2018}.  This signal results from a cluster moving transverse to the line of sight, causing CMB photons to see a different potential on their way into the cluster as on their way out.  Unlike the rkSZ, the moving lens signal will be correlated with the direction of the transverse velocity of the cluster on the sky.  One could imagine that the direction of transverse motion being orthogonal to the cluster rotation axis, which would produce a moving lens dipole aligned with the rkSZ dipole.  However, the ordering of the hot and cold sides of the rkSZ signal relative to transverse motion will be observer dependent: viewed from one side of the cluster, the hot part of the rkSZ signal will be towards the direction of transverse motion, while viewed from a different side, the cold part of the rkSZ signal will be towards the direction of transverse motion.  Consequently, even if the cluster rotation axis and transverse velocity are always orthogonal, any dipole signal due to the moving lens effect should average out in a rotation-oriented stack across multiple clusters.

We have argued that the rkSZ is nearly unique in causing a cluster-rotation-correlated dipole signal in the CMB. One possible exception would be rotational Doppler boosting of the infrared emission from co-rotating galaxies. However, at the frequencies relevant to the \smicanosz{} maps, and for the very low redshift clusters ($z<0.1$) considered here, this is expected to be highly subdominant to SZ signals. Similar arguments apply to any possible co-rotating radio sources (with which the presence of a spatially diffuse signal distributed over many clusters also appears inconsistent). 

While it is otherwise difficult to imagine scenarios where a dipole signal in the CMB that is not the rkSZ is correlated with cluster rotation, there are several potential sources of noise that are expected to correlate with cluster locations, and could degrade the constraints presented here.  For instance, any submillimeter emission from the cluster (such as from dusty galaxies or radio sources) could introduce additional noise.  These noise sources are not taken into account in our noise model, which only includes noise that is uncorrelated with the cluster positions.  While much of the fluctuating emission is expected to be symmetric around the cluster centre and hence not to contribute strongly to the dipolar rkSZ statistic and is in any case expected to be small at low redshifts, we postpone a more careful consideration of these correlated noise sources to future work.  

Similarly, the kSZ due to the overall motion of the cluster is also a potential noise source. The amplitude of this signal (and its sign) will vary depending on the relative motion of the cluster towards or away from the observer. For a spherical cluster, this kSZ signal is a monopole, and will therefore not introduce additional noise to the asymmetry analysis or model fitting results presented here.  Of course, real clusters are not perfect spheres, and asymmetry in the non-rotational kSZ could contribute noise to our measurements.

There are also several modeling approximations we have made that could in principle bias our parameter constraints.  For instance, \citet{Baldi:2018} have shown that the solid body rotation model adopted here does not perfectly match their simulated rkSZ signals.  Furthermore, we have for simplicity treated the rotation of the galaxy members as a perfect proxy for the rotation of the gas.  A more sophisticated model could allow for differences between the gas rotation and galaxy member rotation.  For these reasons, we again caution against over-interpreting the parameter constraints reported here.  However, while these choices may impact the precise values of the constrained parameters, they should not yield a false detection or non-detection of the rkSZ effect.  We postpone a more careful attempt at modeling the signal to future work.

Finally, we also repeat our analysis with the more conservative cluster selection discussed in \S\ref{sec:rot_data}.  We find similar results in this case, albeit with somewhat lower statistical significance.  Given that the conservative selection includes only six clusters, one might not expect to see {\it any} signal in the analysis with this selection.  Our results suggest that the clusters in the conservative selection contribute a significant fraction of the signal-to-noise of the fiducial analysis.  This is not surprising, since the conservative selection would be expected to yield clusters with higher signal.  We note, though, that no individual cluster shows a significant rkSZ detection.

\subsection{Future work}

The prospects for future rkSZ measurements are exciting.  The present analysis with data, and the predictions of \citet{Chluba:2002} and \citet{Baldi:2018}, find a rkSZ signal amplitude of the order 10 to 40~$\mu$K for massive rotating clusters. Ongoing and upcoming CMB experiments like Advanced ACTPol \citep{Henderson:2016}, SPT-3G \citep{Benson:2014}, and the Simons Observatory \citep{SimonsObs2018}  will have roughly few-$\mu$K-arcmin noise over large fractions of the sky.  Measurements of the rkSZ with these data sets that make use of cluster rotation priors will likely yield high signal-to-noise detections.

With the increased sensitivity of future experiments, one could also consider measuring the rkSZ without relying on rotation priors.  To do this, one must be able to discriminate the rkSZ signal from other dipole signals like lensing and the moving lens effect.  Contamination from the lensing signal could be reduced by using information about the large scale CMB gradient at the cluster location.  Similarly, the impact of the moving lens effect could be reduced by using information about the large scale velocity field.  We leave careful discussion of these analysis challenges to future work.

As signal-to-noise ratios on detections of rkSZ increase, it is worth further considering the potential utility of such a signal. The uses of the rkSZ signal broadly divide into astrophysical and cosmological applications. 

Since the rkSZ provides direct insight into the rotational velocity of cluster gas, there may be several astrophysical uses. For example, the gas clusters acquire by tidal stripping of infalling galaxies should have similar specific angular momentum to the cluster member galaxies; by comparing the angular momenta of the gas and the galaxies, knowledge of gas velocities can provide insight into the formation history of clusters. Similarly, improved knowledge of the gas rotation velocity could provide insight into key quantities describing the ICM, and contribute to our understanding of non-thermal pressure support.

In addition, knowledge of the rkSZ could provide new insights into cosmology. In principle, the rotation velocities probed by rkSZ could be related to the tidal tensor and the mass distribution at earlier times, contributing to a reconstruction of the early-time matter field at nearby positions; this could enable powerful constraints on structure growth and related parameters such as neutrino mass (potentially via sample-variance cancellation techniques). While it is currently not clear whether the initial density field can be reconstructed well from rotation velocities (given complications from baryonic or astrophysical effects), the fact that total angular momentum may be approximately conserved within clusters could make this problem more tractable. 

Though there are thus several scientific applications of the rkSZ effect, we defer a detailed consideration of possible opportunities to future work.

\section{Acknowledgements}

We thank James Aguirre, Nicholas Battaglia, Anthony Challinor, Colin Hill, and David Spergel for useful discussions related to this project.

EJB is partially supported by the US
Department of Energy grant DE-SC0007901.  BDS acknowledges support from an Isaac Newton Trust Early Career Grant and an STFC Ernest Rutherford Fellowship

\bibliography{thebib}

\end{document}